\documentclass[twocolumn,showpacs,preprintnumbers,amsmath,amssymb]{revtex4}

\usepackage{graphicx}
\usepackage{dcolumn}
\usepackage{bm}
\usepackage{latexsym}
\usepackage{amsfonts}
\usepackage{amssymb}
\usepackage{amsmath}

\begin{document}

\preprint{KUNS-1930}

\title{ A Unified View of RS Braneworlds }

\author{Sugumi Kanno}
\email{sugumi@tap.scphys.kyoto-u.ac.jp}
\author{Jiro Soda}
\email{jiro@tap.scphys.kyoto-u.ac.jp}
\affiliation{
 Department of Physics,  Kyoto University, Kyoto 606-8501, Japan
}%

\date{\today}

\begin{abstract}
  There are various different descriptions of Randall-Sundrum (RS) braneworlds.
  Here we present a unified view of the braneworld based on
  the gradient expansion approach. In the case of the single-brane model,
  we reveal the relation between
   the geometrical and the AdS/CFT approach. It turns out
    that the high energy and the Weyl term corrections found in the geometrical
    approach merge into the CFT matter correction found in the 
    AdS/CFT approach.   We also clarify the role of the radion
  in the two-brane system. It is shown that the radion transforms 
  the Einstein theory with
  Weyl correction into the conformally coupled scalar-tensor theory 
  where the radion plays the role of the scalar field. 
\end{abstract}

\pacs{98.80.Cq, 98.80.Hw, 04.50.+h}
\maketitle


\section{Introduction}

  Nowadays, the most promising and seemingly a unique candidate 
  for quantum theory of gravity is the string theory. 
  Remarkably, it can be consistently formulated only in 10 dimensions
  ~\cite{Polchinski}. 
  This fact requires a mechanism to fill the gap between our real world and
 the higher dimensions.  Conventionally, the extra dimensions are considered
to be compactified to a small compact space of the Planck scale.
 However, recent developments of superstring theory invented 
 a new idea, the so-called braneworld. 
 The brane world scenario has been the subject of intensive investigation
  for the past few years. Although there are many braneworld 
 models, there are similarities in those models.
 Hence,  we study a simple toy model constructed by Randall and Sundrum
 as a representative~\cite{RS1,RS2}. 
 
  There are various views of RS braneworlds depending on the approach one uses.
 The purpose of this paper is to unify the various views using the
 low energy gradient expansion method~\cite{KS,ShiKo} and give insights into
 the physics of the braneworld.  

 The organization of this paper is as follows:
 In Sec.II, we summarize various views of the braneworlds obtained by
 different methods. In Sec.III, key questions are presented. 
 In Sec.IV, we explain the gradient expansion method. 
 In Sec.V and VI, the single-brane model and the two-brane model
 are analyzed separately. The final section is devoted to the answers
 to the key questions. 

\section{Various Views of Braneworlds}

  Randall and Sundrum   
  proposed a simple model
   where the four-dimensional brane with tension $\sigma$
   is embedded in the five-dimensional asymptotically anti-de Sitter (AdS) bulk
    with a curvature scale $\ell$.
   This single-brane model is described by the action~\cite{RS2}
\begin{eqnarray}
S&=&{1\over 2\kappa^2}\int d^5 x \sqrt{-g}
	\left( 
	{\cal R}+{12\over \ell^2}
	\right)\nonumber \\ 
	&&\quad-\sigma \int d^4 x \sqrt{-h}
	+\int d^4 x \sqrt{-h} {\cal L}_{\rm matter} \ ,
\end{eqnarray}
where ${\cal R}$ and $\kappa^2  $  are the scalar curvature and 
gravitational constant in five-dimensions, respectively. We  impose 
 $Z_2$ symmetry on this spacetime, with the brane at
 the fixed point. The matter ${\cal L}_{\rm matter}$ is 
 confined to the brane. Throughout this paper,
 $h_{\mu\nu}$ represents the induced metric on the brane.  
 Remarkably, the internal dimension is non-compact in this model.
 Hence, we do not have to care about the stability problem. 
 
 Originally, they proposed the two-brane model as a possible solution
 of the hierarchy problem~\cite{RS1}.
  The action reads
\begin{eqnarray}
S&=&\int d^5 x \sqrt{-g} 
	\left( {\cal R} 
	+{12 \over \ell^2}
	\right) 
	-\sum_{i=\oplus ,\ominus} \sigma_i \int d^4 x \sqrt{-h_i } 
	\nonumber \\
	&&\quad+\sum_{i=\oplus ,\ominus}
	\int d^4 x \sqrt{-h} {\cal L}_{\rm matter}^i
        \ , \label{action-5d}
\end{eqnarray}
where $\oplus$ and $\ominus$ represent the positive and the negative
 tension branes, respectively. 

 Both models are prototypes of other models. Here, we list up
 the various approaches to understand these prototypes. 
  
\subsection{Cosmological Approach}

 The homogeneous cosmology of the single-brane model
 is  the  simplest case to be studied. 
 It is easy to deduce  the effective Friedmann equation as
\begin{eqnarray}
   H^2 = {\kappa^2 \over \ell} \rho + \kappa^4 \rho^2 
                     + {{\cal C} \over a_0^4} \ ,
\end{eqnarray}
where $H$, $a_0$ and $\rho$ are, respectively, the Hubble parameter, 
the scale factor and the total energy density of each brane.
The Newton's constant can be identified as $8\pi G_N = \kappa^2 /\ell$. 
 Here, ${\cal C}$ is a constant of integration associated with the mass of
a black hole in the bulk. 
 This constant ${\cal C}$ is referred to as the dark radiation
 in which the effect of the bulk is encoded.  

 There are two kinds of corrections, the high energy correction $\rho^2$
 and the bulk correction ${\cal C}$ which exists even in the low energy 
 regime. Thus, the deviation from the conventional Einstein theory 
 is expected even in the low energy regime. 

 As to the two-brane model, the same effective Friedmann equation can be
 expected because this equation can be deduced without referring to the 
 bulk equations of motion.  In the two-brane case, however, 
 the meaning of 
 ${\cal C}$ is obscure.    
 
\subsection{Linear Perturbation Approach}

  The other useful method to investigate the braneworld is the linearized
  analysis. 
  
  In the case of the single-brane model,
  it was shown that the gravity is localized on the brane in spite of 
  the noncompact extra dimension. Consequently, it turned out that 
   the conventional linearized Einstein equation 
   approximately holds at  scales large compared with
   the curvature scale $\ell$. It should be stressed that this result
   can be attained by imposing the outgoing boundary conditions. 
  
   In the case of the two-brane model, 
Garriga and Tanaka analyzed linearized gravity
 and have shown that the gravity  on the brane 
behaves as the Brans-Dicke theory at low energy~\cite{GT}.
 Thus, the conventional linearized Einstein equations 
do not hold even on scales large compared with the curvature scale $\ell$ 
in the bulk. Charmousis et al. have clearly identified the Brans-Dicke field 
as the radion mode~\cite{CR}.

\subsection{AdS/CFT Correspondence Approach}
 
 There is a clever approach using the concept of the string theory.
After solving the equations of motion in the bulk with the boundary
 value fixed and substituting the solution $\Phi_{\rm cl}$ 
 into the 5-dimensional Einstein-Hilbert action $S_{\rm 5d}$,
 we obtain the effective action for the boundary field
  $\phi= \Phi_{\rm cl}|_{\rm boundary}$.
 The statement of the AdS/CFT correspondence is that the resultant
 effective action can be equated with the partition functional 
 of some conformally invariant field theory (CFT), namely     
\begin{eqnarray}
   \exp \left[ i S_{\rm 5d} [ \Phi_{\rm cl}] \right] 
  \approx < \exp i \int \phi {\cal O}  >_{\rm CFT} \ , 
\end{eqnarray}
where ${\cal O} $ is the field in CFT.
This action must be defined at the infinity where the conformal symmetry
 exists as the asymptotic symmetry. Hence, there exist infrared divergences
 which must be subtracted by the counter terms. 
 Thus, the correct formula becomes
\begin{eqnarray}
   \exp \left[ i S_{\rm 5d} [ \Phi_{\rm cl}] + i S_{\rm ct} \right] 
  = < \exp i \int \phi {\cal O}  >_{\rm CFT} \ , 
\end{eqnarray}
where we added the counter terms
\begin{eqnarray}
    S_{\rm ct} =  S_{\rm brane} - S_{\rm 4d} 
                     - [~R^2 {\rm terms}~]    \ ,
\end{eqnarray}
where $S_{\rm brane}$ and $S_{4d}$ are the brane action  and
the 4-dimensional Einstein-Hilbert action, respectively.
Here, the higher curvature terms $[R^2 {\rm terms}]$ should be 
understood symbolically.

 In the case of the braneworld, the brane acts as the cutoff.
 Therefore, there is no divergences in the above expressions.
 Hence, we can freely rearrange the terms as follows 
\begin{eqnarray}
S_{\rm 5d}+S_{\rm brane}=S_{\rm 4d}+S_{\rm CFT}+[~R^2 {\rm terms}~]
\end{eqnarray}
This tells us that the brane models can be described as the conventional
 Einstein theory with the cutoff CFT and higher order curvature terms.
In terms of the equations of motion, the AdS/CFT correspondence reads
\begin{equation}
\overset{(4)}{G}{}_{\mu\nu} 
	={\kappa^2\over\ell}\left( T_{\mu\nu}+T^{\rm CFT}_{\mu\nu} \right)
	+[~R^2 {\rm terms}~]   \ ,
\end{equation}
where the $R^2$ terms represent the higher order curvature terms and 
 $T^{CFT}_{\mu\nu}$ denotes the energy-momentum tensor of the cutoff version 
 of  conformal field theory.

\subsection{Geometrical Approach}

Here, let us review the geometrical approach~\cite{ShiMaSa}.  
In the  Gaussian normal coordinate system: 
\begin{eqnarray}
ds^2=dy^2+g_{\mu\nu}(y,x^{\mu} )dx^{\mu}dx^{\nu}  \ ,
\label{GN}
\end{eqnarray}
we can write  the 5-dimensional Einstein tensor $\overset{(5)}{G}{}_{\mu\nu}$
in terms of the 4-dimensional Einstein tensor $\overset{(4)}{G}{}_{\mu\nu}$
 and the extrinsic curvature as 
\begin{eqnarray}
\overset{(5)}{G}{}_{\mu\nu}&=& 
	\overset{(4)}{G}{}_{\mu\nu} + K_{\mu\nu,y} - g_{\mu\nu} K_{,y}
	-KK_{\mu\nu}+2K_{\mu\lambda}K^{\lambda}{}_{\nu} \nonumber \\
	&&\qquad+{1\over 2}g_{\mu\nu} 
	\left(K^2+K^\alpha{}_\beta K^\beta{}_\alpha \right)  \nonumber\\
	&=&{6\over\ell^2}g_{\mu\nu}\ ,
	\label{Gauss}
\end{eqnarray}
where  we have introduced the extrinsic curvature 
\begin{eqnarray}
K_{\mu\nu} = - {1\over 2} g_{\mu\nu ,y}  \ ,
\end{eqnarray}
and the last equality comes from the 5-dimensional Einstein equations.
Taking into account the $Z_2$ symmetry,   we also obtain the junction 
 condition 
\begin{eqnarray}
\left[K^\mu{}_\nu-\delta^\mu_\nu K\right]\Big|_{y=0} 
    = {\kappa^2 \over 2}\left(-\sigma\delta^\mu_\nu 
	+T^\mu{}_\nu \right)     \ .
\end{eqnarray}
Here, $T_{\mu\nu}$ represents the 
energy-momentum tensor of the matter. 
Evaluating Eq.~(\ref{Gauss}) at the brane and substituting the 
junction condition 
into it, we have the ``effective" equations of motion 
\begin{eqnarray}
\overset{(4)}{G}{}_{\mu\nu} 
	={\kappa^2\over\ell}T_{\mu\nu}+\kappa^4\pi_{\mu\nu} 
	-E_{\mu\nu}
	\label{ShiMaSa}
\end{eqnarray}
where
\begin{eqnarray}                   
 \pi_{\mu\nu}&=&-{1\over 4}T_\mu{}^\lambda T_{\lambda\nu} 
  +{1\over 12}TT_{\mu\nu} \nonumber\\
&&\qquad\qquad\quad  + {1\over 8} g_{\mu\nu} \left( 
   T^{\alpha\beta}T_{\alpha\beta} -{1\over 3} T^2 \right) \nonumber \\
   E_{\mu\nu} &=& C_{y\mu y \nu} |_{y=0}  \nonumber \ ,
\end{eqnarray}
Here $C_{y\mu y \nu}$ is the Weyl tensor.
We assumed the relation
\begin{eqnarray}
\kappa^2 \sigma = \frac{6}{\ell}
\label{relation}
\end{eqnarray}
so that the effective
 cosmological constant vanishes. 

The geometrical approach is useful to classify possible corrections
 to the conventional Einstein equations. One defect of this approach
 is the fact that the projected Weyl tensor
 can not be determined without solving the equations in the bulk.

\section{Key Questions}

As a landmark, we set a sequence of questions.
 We consider the single-brane model and two-brane model, separately.

\subsection{Single-brane model}

\noindent
{ \bf  Is the Einstein theory recovered even in the non-linear regime?}\\

In the case of the linear theory, it is known that the conventional
 Einstein theory is recovered at low energy. 
 
 On the other hand, the cosmological consideration suggests the deviation
 from the conventional Friedmann equation even in the low energy regime. 
 This is due to the dark radiation term. 
 
 Therefore, we need to clarify
 this discrepancy.  

\vskip 0.5cm
\noindent
{\bf How does the AdS/CFT come into the braneworld?}\\

 It was argued that the cutoff CFT comes into the braneworld.
 However, no one knows what is the cutoff CFT.
 It is a vague concept at least from the point of view of the classical
 gravity. 
  Moreover, it should be noted that the AdS/CFT  correspondence is a specific 
  conjecture. Indeed, originally, Maldacena conjectured that 
  the supergravity on $AdS_5 \times S^5$ is dual to the four-dimensional 
 ${\cal N}=4$ super Yang-Mills theory~\cite{malda}. 
 Nevertheless, the AdS/CFT correspondence seems to
 be related to the brane world model as has been  demonstrated 
 by several people~\cite{ads}.

 Hence, it is important to reveal the role of the AdS/CFT correspondence
 starting from the 5-dimensional general relativity.

\vskip 0.5cm
\noindent
{\bf How are the AdS/CFT and geometrical approach related?}\\

The geometrical approach gives 
\begin{eqnarray*}
 \overset{(4)}{G}{}_{\mu\nu} = {\kappa^2 \over \ell} T_{\mu\nu} 
               + \kappa^4 \pi_{\mu\nu} 
                   - E_{\mu\nu}  \ .
\end{eqnarray*}
 On the other hand, the AdS/CFT correspondence yields
\begin{equation}
   \overset{(4)}{G}{}_{\mu\nu} 
   = {\kappa^2 \over \ell} \left( T_{\mu\nu} + T^{\rm CFT}_{\mu\nu} \right)
                   +[~R^2 {\rm terms}~]   \ .\nonumber
\end{equation}
An apparent difference is remarkable. 

It is an interesting issue to 
 clarify how  these two descriptions are related.  
  Shiromizu and Ida tried to understand the AdS/CFT correspondence 
  from the  geometrical point of view~\cite{SI}.
   They argued that $\pi^\mu_{\ \mu}$ 
  corresponds to the trace anomaly of the cutoff CFT on the brane. 
  However, this  result is rather  paradoxical because there exists 
  no trace anomaly 
  in an odd dimensional  brane although  $\pi^\mu_{\ \mu}$ exists even in that 
  case.  Thus,  the more precise relation between the 
   geometrical and the AdS/CFT approaches remains to be understood. 
  
    Since both the geometrical and AdS/CFT approaches seem 
  to have their own merit, 
  it would be beneficial to understand the mutual relationship.

\subsection{Two-brane model}

\noindent
{\bf How is the geometrical approach consistent with the Brans-Dicke
 picture?}\\

Irrespective of the existence of other branes, the geometric approach gives
 the effective equations (\ref{ShiMaSa}). 
The effect of the bulk geometry comes into
the brane world only through $E_{\mu\nu}$. 
In this picture, the two-brane system can be regarded as the Einstein
 theory with some corrections due to the Weyl tensor in the bulk.

On the other hand,  the linearized gravity  on the brane 
behaves as the Brans-Dicke 
theory  on scales large compared with the curvature scale $\ell$ 
in the bulk~\cite{GT}. 
 Therefore, the conventional linearized Einstein equations 
 do not hold at low energy. 

In the geometrical approach, no radion appears. While, from the linear
 analysis, it turns out that 
 the system can be described by the Brans-Dicke theory where
 the extra scalar field is nothing but the radion.  
How can we reconcile these seemingly incompatible pictures? 

\vskip 0.5cm
\noindent
{\bf What replaces  the AdS/CFT correspondence  
in the two-brane model?}\\

In the single-brane model, 
there are continuum Kaluza-Klein (KK)-spectrum around the zero mode.
 They induce the CFT matter in the 4-dimensional effective action.
 
 In the two-brane system, the spectrum become discrete.
 Hence, we can not expect CFT matter on the brane. Nevertheless, 
 KK-modes exist and  affect the physics on the brane. 
 
 So, it is still interesting
 to know what kind of 4-dimensional theory mimics the effect of the KK-modes.

\section{Gradient Expansion Method}

 Our claim in this paper is that the gradient expansion
 method gives the answers to all of the questions presented in the 
 previous section. Here, we give the formalism developed by us~
\cite{KS}.  

We use the Gaussian normal coordinate system (\ref{GN}) to describe
the geometry of the brane world. 
Note that the brane is located at $y=0$ in this coordinate system. 
Decomposing the extrinsic curvature into the traceless part and the trace part
\begin{equation}
  K_{\mu\nu} 
    = \Sigma_{\mu\nu }+ {1\over 4} h_{\mu\nu} K  \ , \quad
    K = - {\partial \over \partial y}\log \sqrt{-g}    \  , 
\end{equation}
we obtain the basic equations which hold in the bulk;
\begin{eqnarray}
&&\Sigma^\mu{}_{\nu ,y}-K\Sigma^\mu{}_{\nu} 
	=-\left[
	\overset{(4)}{R}{}^\mu{}_\nu 
	-{1\over 4} \delta^\mu_\nu \overset{(4)}{R} 
        \right]
        \label{munu-trc} \ ,     \\
&&{3\over 4}K^2-\Sigma^\alpha{}_{\beta}\Sigma^\beta{}_{\alpha} 
	=\left[~
	\overset{(4)}{R}
	~\right] 
	+{12\over\ell^2}
	\label{munu-trclss} \ ,  \\
&&K_{, y}-{1\over 4}K^2-\Sigma^{\alpha\beta}\Sigma_{\alpha\beta} 
	=-{4\over\ell^2}
	\label{yy} \ ,  \\
&&\nabla_\lambda\Sigma_{\mu}{}^{\lambda}  
	-{3\over 4}\nabla_\mu K = 0
	\label{ymu} \ ,
\end{eqnarray}
where $\overset{(4)}{R}{}^\mu{}_\nu$ 
is the curvature on the brane and $\nabla_\mu $ denotes the
 covariant derivative with respect to the metric $g_{\mu\nu}$.
 One also have the junction condition 
\begin{equation}
\left[
	K^\mu{}_\nu -\delta^\mu_\nu K
	\right]
	\Big|_{y=0} 
	={\kappa^2\over 2}
	\left(
	-\sigma\delta^\mu_\nu+T^\mu{}_\nu
	\right) \ .
\end{equation}
Recall that we are considering the $Z_2$ symmetric spacetime. 

The problem now is separated into two parts. First, we will solve
 the bulk equations of motion with the Dirichlet boundary condition
 at the brane, $g_{\mu\nu} (y=0 ,x^\mu ) = h_{\mu\nu} (x^\mu ) $.
 After that, the junction condition will be imposed at the brane.
 As it is the condition for the induced metric $h_{\mu\nu}$, it
 is naturally interpreted as the effective equations of motion
 for  gravity on the brane.
 
Along the normal coordinate $y$, the metric varies with a characteristic length
 scale $\ell$; $ g_{\mu\nu ,y} \sim g_{\mu\nu} /\ell$.  Denote the 
characteristic 
 length scale of the curvature fluctuation on the brane as $L$; then we have
  $ R \sim g_{\mu\nu} / L^2 $. For the reader's reference, let us take 
  $\ell=1$ mm, for example. Then, the relation (\ref{relation}) give 
  the scale, 
  $\kappa^2 = (10^8 \ {\rm GeV})^{-3}$ and $\sigma = 1 \ {\rm TeV}^4$.  
  In this paper, we will consider the low energy
  regime in the sense that the energy density of matter, $\rho$, 
  on the brane is smaller than the brane tension, \i.e., $\rho /\sigma \ll 1$. 
  In this regime, a simple dimensional analysis
\begin{equation}
  {\rho \over \sigma} 
  \sim {\ell {\kappa^2 \over \ell} \rho \over \kappa^2 \sigma}
    \sim \left({\ell\over L}\right)^2 \ll 1 
\end{equation}
implies that the curvature on the brane can be neglected compared with the
 extrinsic curvature at low energy. Here, we have used the relation 
 (\ref{relation})
 and Einstein's equation on the brane, 
 $R\sim g_{\mu\nu}/L^2 \sim \kappa^2\rho/\ell$.
 Thus, the anti-Newtonian or gradient expansion method used in the cosmological
 context is applicable to our problem~\cite{tomita}.

At zeroth order,  we can neglect the curvature term. Then we have
\begin{eqnarray}
&&\overset{(0)}{\Sigma}{}^{\mu}{}_{\nu , y}
	-\overset{(0)}{K}\overset{(0)}{\Sigma}{}^{\mu}{}_{\nu} 
	=0
	\label{0:munu}     \ ,  \\
&&{3\over 4}\overset{(0)}{K}{}^{2} 
	-\overset{(0)}{\Sigma}{}^{\alpha}{}_{\beta} 
	\overset{(0)}{\Sigma}{}^{\beta}{}_{\alpha} 
	={12\over \ell^2}   \ ,  \\
&&\overset{(0)}{K}{}_{, y} -{1\over 4} \overset{(0)}{K}{}^{2} 
	-\overset{(0)}{\Sigma}{}^{\alpha \beta} 
	\overset{(0)}{\Sigma}{}_{\alpha \beta} 
	=-{4\over \ell^2}     \ ,   \\
&&\nabla_\lambda\overset{(0)}{\Sigma}{}^{\lambda}{}_{\mu }   
	-{3\over 4}\nabla_\mu \overset{(0)}{K} = 0
	\label{0:ymu} \ .
\end{eqnarray}

 Equation (\ref{0:munu}) can be readily integrated into
\begin{equation}
    \overset{(0)}{\Sigma}{}^{\mu}{}_{\nu} 
    = {C^\mu{}_{\nu} (x^\mu) \over \sqrt{-g} } 
                                 \ , \quad C^\mu{}_{\mu} =0  \ ,
\end{equation}
where $C^\mu{}_{\nu}$ is the constant of integration. 
 Equation (\ref{0:ymu}) also requires $C^\mu{}_{\nu |\mu} =0$.
 If it could exist, it would represent a radiation like fluid 
 on the brane and hence  a strongly anisotropic universe. 
 In fact, as we see soon, this
 term must vanish in order to satisfy the junction condition. 
 Therefore, we simply put $C^\mu{}_\nu =0$, hereafter. 
 Now, it is easy to solve the remaining equations. The result is
\begin{equation}
    \overset{(0)}{K} = {4\over \ell}    \ .
\end{equation}
Using the definition of the extrinsic curvature
\begin{equation}
     \overset{(0)}{K}{}_{\mu\nu} 
     = - {1\over 2} {\partial \over \partial y} 
                               \overset{(0)}{g}{}_{\mu\nu}     \   ,
\end{equation}
we get the zeroth order metric  as
\begin{equation}
 ds^2 = dy^2 +  a^2 (y) h_{\mu\nu}(x^\mu ) dx^\mu dx^\nu  \ , \quad
    a(y)  = e^{-2{y\over \ell}}    \ ,
\end{equation}
where  the tensor $h_{\mu\nu}$ is  the induced metric on the brane. 

From the zeroth order solution, we obtain
\begin{equation}
   \left[ \overset{(0)}{K}{}^{\mu}{}_{\nu} 
   - \delta^\mu_\nu \overset{(0)}{K} \right] \Bigg|_{y=0}
    = -{3 \over \ell} \delta^\mu_\nu
    = - {\kappa^2 \over 2} \sigma \delta^\mu_\nu  \ .
\end{equation}
Then we get the well known relation $\kappa^2 \sigma = 6/\ell$.
Here, we will assume that this relation holds exactly. 
 It is apparent that $C^\mu{}_{\nu}$ is not allowed to exist.

The iteration scheme consists in writing the metric $g_{\mu\nu}$
 as a sum of local tensors built out of the induced metric on the
 brane, the number of gradients increasing with the order. 
Hence, we will seek the metric as a perturbative series
\begin{eqnarray}
     g_{\mu\nu} (y,x^\mu ) &=&
  a^2 (y) \left[ h_{\mu\nu} (x^\mu)  + \overset{(1)}{g}{}_{\mu\nu} (y,x^\mu)
      \right. \nonumber\\
  && \left.  \qquad\qquad \quad
  + \overset{(2)}{g}{}_{\mu\nu} (y, x^\mu ) + \cdots  \right]  \ , 
\end{eqnarray}
where $a^2 (y) $ is extracted 
 and we put the Dirichlet boundary condition
\begin{eqnarray}  
     \overset{(i)}{g}{}_{\mu\nu} (y=0 ,x^\mu ) =  0    \ ,
\end{eqnarray}
so that $g_{\mu\nu} (y=0, x) =  h_{\mu\nu} (x)$ holds at the brane. 
Other quantities can be also expanded as
\begin{eqnarray}
K^\mu{}_{\nu}&=&{1\over\ell}
	\delta^{\mu}_{\nu}
        +\overset{(1)}{K}{}^{\mu}{}_{\nu}
	+\overset{(2)}{K}{}^{\mu}{}_{\nu}+\cdots  \nonumber\\
\Sigma^\mu{}_{\nu}
	&=&\qquad
	+\overset{(1)}{\Sigma}{}^{\mu}{}_{\nu}
	+\overset{(2)}{\Sigma}{}^{\mu}{}_{\nu} + \cdots          \ .
\end{eqnarray}
In our scheme,
 in contrast to the AdS/CFT correspondence where the Dirichlet boundary 
 condition is imposed at infinity, we impose it
  at the finite point $y=0$, the location of the brane. 
 Furthermore, we carefully consider the
  constants of integration, i.e., homogeneous solutions. These 
 homogeneous solutions are  ignored in the calculation of AdS/CFT 
 correspondence.  However, they play the important role in the
 braneworld.

\section{Single brane model (RS2)}

\subsection{Einstein Gravity at Lowest Order}

The next order solutions are obtained by taking into account the 
terms neglected at zeroth order. 
At  first order,  Eqs.~(\ref{munu-trc}) - (\ref{ymu}) become
\begin{eqnarray}
&&\overset{(1)}{\Sigma}{}^{\mu}{}_{\nu , y} 
	-{4\over\ell} \overset{(1)}{\Sigma}{}^{\mu}{}_{\nu} 
	=-\left[\overset{(4)}{R}{}^\mu{}_\nu 
	-{1\over 4} \delta^\mu_\nu\overset{(4)}{R}\right]^{(1)}
	\label{1:munu} \ , \\
&&{6 \over\ell} \overset{(1)}{K}  = \left[~ \overset{(4)}{R}
	~\right]^{(1)}   \  ,\\
&&\overset{(1)}{K}{}_{, y} -{2\over \ell} \overset{(1)}{K} = 0   \ ,\\
&&\overset{(1)}{\Sigma}{}_{\mu}{}^{\lambda}{}_{|\lambda}  
	-{3\over 4}\overset{(1)}{K}{}_{|\mu} = 0 \ .
\end{eqnarray}
where the superscript $(1)$ represents the order of the derivative expansion
 and $|$ denotes the covariant derivative with respect to
 the metric $h_{\mu\nu}$.
Here, $[\overset{(4)}{R}{}^\mu{}_\nu ]^{(1)} $ 
means that the curvature is approximated by
 taking the Ricci tensor of $a^2 h_{\mu\nu} $ in place of 
 $\overset{(4)}{R}{}^{\mu}{}_{\nu}$. 
 It is also convenient to write it in terms of the Ricci
  tensor of $h_{\mu\nu}$, denoted $R^\mu{}_\nu (h)$.
 
Substituting the zeroth order metric into $\overset{(4)}{R}$, we obtain
\begin{equation}
\overset{(1)}{K} = {\ell\over 6a^2} R(h)
\label{1:trc} \ .
\end{equation}
Hereafter, we omit the argument of the curvature for simplicity. 
Simple integration of Eq.~(\ref{1:munu}) also gives the traceless part
 of the extrinsic curvature as
\begin{equation}
\overset{(1)}{\Sigma}{}^{\mu}{}_{\nu}={\ell\over 2a^2}
	(R^\mu{}_{\nu}-{1\over 4}\delta^\mu_\nu R)  
	+{\chi^{\mu}{}_{\nu}(x)\over a^4}
	\label{1:trclss}  \ ,
\end{equation}
where the homogeneous solution satisfies the constraints 
\begin{eqnarray}
\chi^{\mu}{}_{\mu}=0   \ , \quad \chi^{\mu}{}_{\nu|\mu}=0 \ .
\label{TT}
\end{eqnarray}
 As we see later, this term  corresponds to  dark 
 radiation at this order.  
The metric can be obtained as
\begin{eqnarray}
  \overset{(1)}{g}{}_{\mu\nu} &=&  -{\ell^2 \over 2} 
  \left( {1\over a^2}-1 \right) 
    \left( R_{\mu\nu}  - {1\over 6} h_{\mu\nu} R \right)
    \nonumber \\ 
  && \quad  -{\ell \over 2}\left( {1\over a^4} -1 \right) \chi_{\mu \nu} \ ,
\end{eqnarray}
where we have imposed the boundary condition, 
$\overset{(1)}{g}{}_{\mu\nu} (y=0, x^\mu ) =0 $.

Let us focus on the role of  $\chi^\mu{}_{\nu}$ in this part.
 At this order, the junction condition can be written as
\begin{eqnarray}
&&\left[~\overset{(1)}{K}{}^{\mu}{}_{\nu} 
	-\delta^\mu_\nu\overset{(1)}{K}~\right] \Bigg|_{y=0}  \nonumber \\
&&\qquad
	={\ell\over 2}\left(
	R^\mu{}_{\nu}-{1\over 2}\delta^\mu_\nu R
	\right)
	+\chi^\mu{}_{\nu}
	={\kappa^2\over 2}T^\mu{}_{\nu}  \ .
\end{eqnarray}
Using the solutions Eqs.~(\ref{1:trc}), (\ref{1:trclss}) and the formula
\begin{equation}
E^\mu{}_{\nu}=K^\mu{}_{\nu ,y}-\delta^\mu_\nu K_{,y}
	-K^\mu{}_{\lambda}K^\lambda{}_{\nu}  
	+\delta^\mu_\nu K^\alpha{}_{\beta}K^\beta{}_{\alpha}
	-{3\over\ell^2}\delta^\mu_\nu   \ ,
\end{equation}
we  calculate the projective Weyl tensor as 
\begin{eqnarray}
\overset{(1)}{E}{}^{\mu}{}_{\nu}={2 \over\ell}\chi^\mu{}_{\nu} \ .
\end{eqnarray}
Then we obtain the effective Einstein equation
\begin{equation}
R^\mu{}_{\nu}-{1\over 2}\delta^\mu_\nu R  
	={\kappa^2\over\ell}T^\mu{}_{\nu}-\overset{(1)}{E}{}^{\mu}{}_{\nu}
	\label{1:effeq} \ .
\end{equation}
At this order, we do not have the conventional Einstein equations. 
 Recall that  the dark radiation exists even 
 in the low energy regime. Indeed, the low energy effective Friedmann
 equation becomes
\begin{equation}
  H^2 = {\kappa^2 \over 3\ell} \rho + {{\cal C} \over a_0 (t)^4}  \ .
\end{equation}
This equation can be obtained from Eq.~(\ref{1:effeq}) by
 imposing the maximal symmetry on the spatial part of the brane world
  and  the conditions (\ref{TT}).
Hence, we observe that $\chi^\mu{}_{\nu}$ is the generalization of 
the dark radiation found in the cosmological context. 

The nonlocal tensor $\chi_{\mu\nu}$ must be determined by the 
   boundary conditions in the bulk. 
 The natural choice is  asymptotically  AdS boundary condition.
 For this boundary condition,  $\chi_{\mu\nu} =0 $. 
  It is this boundary condition that
  leads to the conventional Einstein theory in linearized gravity. 
Assuming this, we have
\begin{equation}
R^\mu{}_{\nu}-{1\over 2}\delta^\mu_\nu R  
	={\kappa^2\over\ell}T^\mu{}_{\nu}   \ .
\end{equation}
Thus,  Einstein theory is recovered at the leading order!

\subsection{AdS/CFT Emerges}

In this subsection, we do not include the $\chi$ field 
because we have adopted the AdS boundary condition. 
Of course, we have  calculated the second order solutions 
with the contribution of the $\chi$ field.
 It merely adds  extra terms such as $\chi^\mu{}_{\nu}\chi^\nu{}_{\mu}$, etc. 
 
At  second order, the basic equations can be easily deduced. 
Substituting the solution up to  first order into the Ricci tensor
 and picking up the second order quantities, we obtain 
 the solutions at second order. 
The trace part is deduced algebraically as
\begin{eqnarray}
\overset{(2)}{K}
	&=& {\ell^3 \over 8 a^4} \left( R^\alpha{}_{\beta}R^\beta{}_{\alpha} 
	-{2\over 9}R^2 \right) 
	\nonumber \\
	&&\quad
	-{\ell^3 \over 12a^2} 
	\left( R^\alpha{}_{\beta} R^\beta{}_{\alpha} 
	-{1\over 6}R^2 \right)   \ .
\end{eqnarray}
 By integrating the equation for the traceless part, we have
\begin{eqnarray}
\overset{(2)}{\Sigma}{}^{\mu}{}_{\nu}  
	&=&-{\ell^2\over 2}\left( {y\over a^4} 
	+{\ell\over 2a^2} \right)
	{\cal S}^\mu{}_{\nu}  \nonumber \\
	&&-{\ell^3 \over 24 a^2} 
	\left( RR^\mu{}_{\nu} 
	-{1\over 4}\delta^\mu_\nu R^2
	\right) 
	+{\ell^3 \over a^4}t^\mu{}_{\nu}   \ ,
\end{eqnarray}
where ${\cal S}^\mu{}_\nu$ is defined by 
\begin{eqnarray}
&&\delta \int d^4 x \sqrt{-h} {1\over 2} \left[ 
	R^{\alpha\beta}R_{\alpha\beta} -{1\over 3}R^2 \right] 
	\nonumber \\
	&&\hspace{2.5cm}
	=\int d^4 x \sqrt{-h}{\cal S}_{\mu\nu}
	\delta g^{\mu\nu} 
	\label{smunu}\ .
\end{eqnarray}
The tensor ${\cal S}^\mu{}_\nu$ is transverse and traceless,
\begin{equation}
{\cal S}^\mu{}_{\nu |\mu} =0  \ ,\quad {\cal S}^\mu{}_{\mu} = 0   \ .
\end{equation}

The homogeneous solution $t^\mu{}_{\nu}$ must be traceless. 
 Moreover, it must satisfy the momentum constraint. 
 To be more precise, we must solve the constraint equation
\begin{equation}
t^\mu{}_{\nu |\mu}-{1\over 16}R^\alpha{}_{\beta}R^\beta{}_{\alpha|\nu}
	+{1\over 48}RR_{|\nu}-{1\over 24}R_{|\lambda}R^\lambda{}_{\nu}  
	=0 \ .
\end{equation}
As one can see immediately,
 there are ambiguities in integrating this equation. 
Indeed, there are two types of covariant local tensor whose
 divergences vanish:
\begin{eqnarray}
&&\delta\int d^4x\sqrt{-h}{1\over 2}R^{\alpha\beta}R_{\alpha\beta} 
	=\int d^4x\sqrt{-h}{\cal H}_{\mu\nu}\delta g^{\mu\nu}
	\label{hmunu} \ , \\
&&\delta\int d^4x\sqrt{-h}{1\over 2}R^2 
	=\int d^4x\sqrt{-h}{\cal K}_{\mu\nu}\delta g^{\mu\nu} 
	\label{kmunu}\ .
\end{eqnarray}
Notice that ${\cal S}^\mu{}_{\nu}={\cal H}^\mu{}_{\nu}
-{\cal K}^\mu{}_{\nu}/3 $. Hence, only ${\cal S}^\mu{}_\nu$ and
 ${\cal K}^\mu{}_\nu$ are independent. 
 Thanks to the Gauss-Bonnet topological 
 invariant, we do not need to consider the Riemann squared term.  
 In addition to these local tensors, we have to take into account
  the nonlocal tensor $\tau^\mu{}_{\nu}$ 
  with the property $\tau^\mu{}_{\nu|\mu}=0$. Thus, we get
\begin{eqnarray}
t^\mu{}_{\nu} 
	&=&{1\over 32}\delta^\mu_\nu 
	\left( R^\alpha{}_{\beta}R^\beta{}_{\alpha} 
	-{1\over 3}R^2 \right) 
	+{1\over 24}\left( RR^\mu{}_{\nu} 
         -{1\over 4}\delta^\mu_\nu R^2\right) \nonumber\\
	&&\qquad+\tau^\mu{}_{\nu}  
        +\left( \alpha+{1\over 4} \right){\cal S}^\mu{}_{\nu} 
        +{\beta\over 3}{\cal K}^\mu{}_{\nu}  \ , 
\end{eqnarray}
where the constants $\alpha$ and $\beta$ represents the freedom of 
the gravitational wave in the bulk. 
 The condition  $t^\mu{}_{\mu} =0$ leads to
\begin{equation}
\tau^\mu{}_{\mu}  
	=-{1\over 8}\left( R^\alpha{}_{\beta}R^\beta{}_{\alpha} 
        -{1\over 3}R^2 \right)-\beta\Box R      \ .
\end{equation}
This expression is the reminiscent of the trace anomaly of the CFT.
 It is possible to use the result of CFT at this point.  
 For example, we can choose the ${\cal N}=4$ super Yang-Mills theory
 as the conformal matter. In that case, we  simply put $\beta =0$. 
 This is the
 way the AdS/CFT correspondence comes into the brane world scenario. 

Up to the second order, the junction condition gives
\begin{eqnarray}
R^\mu{}_{\nu}-{1\over 2}\delta^\mu_\nu R 
	+2\ell^2\left[ \tau^\mu{}_{\nu} 
	+\alpha{\cal S}^\mu{}_{\nu}
	+{\beta\over 3}{\cal K}^\mu{}_{\nu}\right] 
	={\kappa^2\over\ell}T^\mu{}_{\nu}
	\label{2:effeq} \ .
\end{eqnarray}
 If we define 
\begin{eqnarray}
T_{\mu\nu}^{\rm CFT}= -2 {\ell^3 \over \kappa^2} \tau_{\mu\nu}  \ , 
\end{eqnarray}
we can write
\begin{eqnarray}
\overset{(4)}{G}{}_{\mu\nu} = {\kappa^2 \over \ell} \left( 
     T_{\mu\nu} + T_{\mu\nu}^{\rm CFT} \right) 
       -2 \ell^2  \alpha  {\cal S}_{\mu\nu} 
         - {2 \ell^2 \over 3} \beta {\cal K}_{\mu\nu}  \ .
\end{eqnarray}

Let us try to arrange the terms so as to reveal the geometrical
 meaning of the above equation. 
 We can calculate the projective Weyl  tensor as
\begin{equation}
\overset{(2)}{E}{}^{\mu}{}_{\nu} 
	=\ell^2\left[P^\mu{}_{\nu}+2\tau^\mu{}_{\nu} 
	+2\alpha{\cal S}^\mu{}_{\nu} 
	+{2\over 3}\beta{\cal K}^\mu{}_{\nu} \right]     \ ,
\end{equation}
where
\begin{eqnarray}
P^\mu{}_{\nu }&=&-{1\over 4}R^\mu{}_{\lambda}R^\lambda{}_{\nu}  
	+{1\over 6}R R^\mu{}_{\nu} \nonumber \\
	&&\qquad+{1\over 8}\delta^\mu_\nu R^\alpha{}_{\beta}R^\beta{}_{\alpha}
	-{1\over 16}\delta^\mu_{\nu}R^2  \ .
\end{eqnarray}
Substituting this expression into Eq.~(\ref{2:effeq}) yields our main result
\begin{equation}
\overset{(4)}{G}{}_{\mu\nu}={\kappa^2\over\ell}T_{\mu\nu}   
	+\ell^2P_{\mu\nu}-\overset{(2)}{E}{}_{\mu\nu}
	\label{2:effeq2} \ .
\end{equation}
Notice that $E^\mu{}_{\nu}$ contains the nonlocal part and the free 
parameters $\alpha$ and $\beta$. 
 On the other hand, $P^\mu{}_{\nu}$ is determined locally. 
 One can see the relationship in a more transparent way.  
Within the accuracy we are considering,  
we can get $ P^\mu{}_{\nu }=\pi^\mu{}_{\nu}$ using the lowest order 
equation 
$R^\mu{}_{\nu}={\kappa^2 /\ell}(T^\mu{}_{\nu}-1/2\delta^\mu_\nu T) $.
 Hence, we can rewrite Eq.~(\ref{2:effeq2}) as 
\begin{equation}
\overset{(4)}{G}{}_{\mu\nu}={\kappa^2\over\ell}
	T_{\mu\nu}+\kappa^4\pi_{\mu\nu} 
	-\overset{(2)}{E}{}_{\mu\nu}
	\label{2:SMS} \ .
\end{equation}
 Now, the similarity between Eqs.~(\ref{ShiMaSa}) and (\ref{2:SMS}) is 
 apparent. 
 Thus we get an explicit relation between the geometrical approach and the
 AdS/CFT approach.
However, we note that our Eq.~(\ref{2:SMS}) is a closed system
 of equations provided that the specific conformal field theory
 is chosen.

Now we can read off the effective action as
\begin{eqnarray}
S_{\rm eff}&=&\frac{\ell}{2\kappa^2}\int d^4x\sqrt{-h}~R
	+S_{\rm matter}+S_{\rm CFT}  \nonumber \\
	&&\quad
	+\frac{\alpha\ell^2}{2\kappa^2}
	\int d^4x\sqrt{-h}
	\left[R^{\mu\nu}R_{\mu\nu}
	-{1\over 3}R^2 \right] \nonumber \\
	&&\quad
	+\frac{\beta\ell^2}{6\kappa^2}\int d^4x\sqrt{-h}~R^2  \ ,
\end{eqnarray}
where we have used the relations Eqs.~(\ref{smunu}), (\ref{hmunu}) 
and (\ref{kmunu}) and we denoted the 
nonlocal effective action constructed from $\tau^\mu{}_{\nu}$ as $S_{\rm CFT}$.

\section{Two-brane model (RS1)}

\subsection{Scalar-Tensor Theory Emerges}

 We consider  the two-brane system in this section.
 Without matter on the branes, we have the relation
$g^{\ominus\hbox{-}\rm brane}_{\mu\nu}
=e^{-2d/\ell}g^{\oplus\hbox{-}\rm brane}
\equiv \Omega^2 g^{\oplus\hbox{-}\rm brane} $
where $d$ is the distance between the two branes.
Although $\Omega$ is constant for vacuum branes, it becomes the
function of the 4-dimensional coordinates if we put the matter on the brane.

Adding the energy momentum tensor to each of the two branes,
and allowing deviations from the pure AdS$_5$ bulk, the
effective (non-local) Einstein equations on the branes at low
energies take the form\cite{KS, ShiMaSa},
\begin{eqnarray}
G^\mu{}_{\nu} (h )
	&=&{\kappa^2\over\ell}\overset{\oplus}{T}{}^{\mu}{}_{\nu} 
	-{2\over \ell}\chi^\mu{}_{\nu} \,,
\label{A:einstein}
\\
G^\mu{}_{\nu}(f)
&=&-{\kappa^2\over\ell}\overset{\ominus}{T}{}^{\mu}{}_{\nu} 
	-{2\over\ell}{\chi^\mu{}_{\nu}\over\Omega^4} \ .
\label{B:einstein}
\end{eqnarray}
where $h_{\mu\nu}=g^{\oplus\hbox{-}{\rm brane}}_{\mu\nu}$, 
$f_{\mu\nu}=g^{\ominus\hbox{-}{\rm brane}}_{\mu\nu}=\Omega^2h_{\mu\nu}$
and the terms proportional to $\chi_{\mu\nu}$ are 5-dimensional
Weyl tensor contributions which describe
the non-local 5-dimensional effect.
Although Eqs.~(\ref{A:einstein}) and (\ref{B:einstein})
are non-local individually, with undetermined $\chi_{\mu\nu}$,
one can combine both equations to reduce them to local equations
for each brane. Since $\chi_{\mu\nu}$ appears only algebraically,
one can easily eliminate $\chi_{\mu\nu}$ from Eqs.~(\ref{A:einstein}) 
and (\ref{B:einstein}). 
Defining a new field $\Psi = 1-\Omega^2$,  we find 
\begin{eqnarray}
\hspace{-3mm}
G^\mu{}_{\nu}(h)&=&{\kappa^2 \over \ell \Psi } 
	\overset{\oplus}{T}{}^{\mu}{}_{\nu}
	+{\kappa^2 (1-\Psi )^2 \over \ell\Psi } 
	\overset{\ominus}{T}{}^{\mu}{}_{\nu} \nonumber \\
&&	+{1\over\Psi}\left(\Psi^{|\mu}{}_{|\nu} 
  	-\delta^\mu_\nu\Psi^{|\alpha}{}_{|\alpha}\right) \nonumber\\
&&	+{3 \over 2 \Psi (1-\Psi )}\left( \Psi^{|\mu}\Psi_{|\nu}
  	- {1\over 2}\delta^\mu_\nu\Psi^{|\alpha}\Psi_{|\alpha} 
  	\right),
  	\label{A:STG1} \\
\hspace{-3mm}
\Box\Psi&=&{\kappa^2\over 3\ell}(1-\Psi )
	\left\{ \overset{\oplus}{T} + (1-\Psi)\overset{\ominus}{T}  
	                     \right\} \nonumber \\
&&	-{1 \over 2 (1-\Psi )} \Psi^{|\mu}\Psi_{|\mu} \ , 
  	\label{A:STG2}  	
\end{eqnarray}
where $|$ denotes the covariant derivative with respect to the metric
$h_{\mu\nu}$. Since $\Omega$ (or equivalently $\Psi$) contains
the information of the distance between the two branes,
we call $\Omega$ (or $\Psi$) the radion.

We can also determine $\chi^{\mu}{}_{\nu}$ by eliminating $G^{\mu}{}_{\nu}$ 
from Eqs.~(\ref{A:einstein}) and (\ref{B:einstein}). Then,
\begin{eqnarray}
\chi^{\mu}{}_{\nu}&=&-{\kappa^2(1-\Psi)\over 2 \Psi} 
	\left( \overset{\oplus}{T}{}^{\mu}{}_{\nu} 
	+(1-\Psi)\overset{\ominus}{T}{}^{\mu}{}_{\nu}\right)  \nonumber\\
&&	-{\ell\over 2\Psi} \left[ \left(  \Psi^{|\mu}{}_{|\nu} 
	-\delta^\mu_\nu  \Psi^{|\alpha}{}_{|\alpha} \right) 
	\right. \nonumber \\
&&	\left.+{3 \over 2(1 -\Psi )} \left( \Psi^{|\mu}  \Psi_{|\nu}
  	-{1\over 2} \delta^\mu_\nu  \Psi^{|\alpha} \Psi_{|\alpha} 
  	\right) \right]   \ .  
  	\label{A:chi}
\end{eqnarray}
 Note that the index of $\overset{\ominus}{T}{}^{\mu}{}_{\nu}$ is to be raised
or lowered by the induced metric on the $\ominus$-brane, $f_{\mu\nu}$.

The effective action for the $\oplus$-brane which gives 
Eqs.~(\ref{A:STG1}) and (\ref{A:STG2}) is
\begin{eqnarray}
S_{\oplus}&=&{\ell \over 2 \kappa^2} \int d^4 x \sqrt{-h} 
	\left[ \Psi R - {3 \over 2(1- \Psi )} 
     	\Psi^{|\alpha} \Psi_{|\alpha} \right] \nonumber\\
    && \!\!\!\!\!\!\!\!\!\!
    	+ \int d^4 x \sqrt{-h} {\cal L}^\oplus 
      	+ \int d^4 x \sqrt{-h} \left(1-\Psi \right)^2 {\cal L}^\ominus  
      	\ .  
      	\label{A:action} 
\end{eqnarray}
It should be stressed that  the radion has the conformal coupling.

\subsection{AdS/CFT in two-brane system?}

 In the two-brane case, it is difficult to proceed to the next order 
 calculations. Hence, we need to invent a new method~\cite{sugumi}. 
 For this purpose, we shall start with the effective Einstein equation obtained 
 by Shromizu, Maeda, and Sasaki 
\begin{eqnarray}
 G_{\mu\nu} = T_{\mu\nu} + \pi_{\mu\nu} - E_{\mu\nu}
\end{eqnarray}
where $\pi_{\mu\nu}$ is the quadratic of energy momentum tensor $T_{\mu\nu}$ and
$E_{\mu\nu}$ represents the  effect of the bulk geometry. 
 Here we have set $8\pi G=1$. 
 This geometrical projection approach can not give a concrete prediction, 
 because we do not know $E_{\mu\nu}$ without solving the equations of motion 
 in the bulk. Fortunately, in the case of the homogeneous cosmology, 
 the property $E^\mu{}_{\mu} =0$ determines the dynamics as 
\begin{eqnarray}
  H^2 = {1  \over 3} \rho + \rho^2 + {{\cal C} \over a_0^4} \ .
\end{eqnarray}
 This reflects
 the interplay between the bulk and the brane dynamics on the brane.  

 What we want to seek is an effective theory which contains the information
 of the bulk as finite number of constant parameters like ${\cal C}$ in the
 homogeneous universe. When we succeed to obtain it, the cosmological 
 perturbation theory can be constructed in a usual way. Although the
 concrete prediction can not be made, qualitative understanding of
 the evolution of the cosmological fluctuations can be obtained.
 This must be useful to make observational predictions.

 In the two-brane system, the mass spectrum is known from the linear
 analysis~\cite{GT}. At low energy, the propagator for the KK mode
 with the mass $m$ can be expanded as 
\begin{eqnarray}
   {-1 \over \Box - m^2} = {1\over m^2}\left[ 1 + {\Box \over m^2}
   + {\Box^2 \over m^4}+ \cdots \right] \ .
\end{eqnarray}
 However, massless modes can not be expanded in this way, hence we must 
 take into account all of the massless modes to construct braneworld
 effective action. 
 It seems legitimate to assume this consideration is valid even 
  in the non-linear regime.  Thus, at low energy,
 the action can be expanded by the local terms with increasing
  orders of derivatives of the metric $g_{\mu\nu}$ 
  and the radion $\Psi$~\cite{KS}.

 Let us illustrate our method using the following action truncated at the second
 order derivatives:
\begin{eqnarray}
 S_{\rm eff} 
 	&=&{1\over 2}  \int d^4 x \sqrt{-g} \left[ 
 	\Psi R -2\Lambda (\Psi)   \right. \nonumber \\
	&&\left.\qquad\qquad\qquad 
	-{\omega (\Psi) \over \Psi} 
	\nabla^\mu\Psi\nabla_\mu\Psi\right]
	\label{1:action} \ ,
\end{eqnarray}
 which is nothing but the scalar-tensor theory with coupling function
 $\omega (\Psi)$ and the potential function $\Lambda (\Psi )$. 
 Note that this is the most general local action which contains 
 up to the second  order derivatives and has the general coordinate invariance.
 It should be stressed that the scalar-tensor theory is, in general,
  not related to the braneworld. However, we know a special type of 
 scalar-tensor theory corresponds to  the low energy 
 braneworld~\cite{KS, ShiKo, wiseman}. 
 Here, we will present a simple derivation of this known fact.  
 
 For the vacuum brane, we can put 
$T_{\mu\nu} + \pi_{\mu\nu} = - \lambda g_{\mu\nu}$. Hence, 
 the geometrical effective equation  reduces to 
\begin{eqnarray}
G_{\mu\nu}=-E_{\mu\nu}-\lambda g_{\mu\nu}
\label{1:SMS}\ .
\end{eqnarray}
 First, we must find $E_{\mu\nu}$. 
The above action (\ref{1:action}) gives the equations of motion for 
the metric as
\begin{eqnarray}
G_{\mu\nu}&=&-{\Lambda \over \Psi} g_{\mu\nu}
	+{1\over \Psi} \left( 
	\nabla_\mu \nabla_\nu \Psi
	-g_{\mu\nu} \Box \Psi \right)
	\nonumber\\
&&	+ {\omega \over \Psi^2} \left(
        \nabla_\mu \Psi \nabla_\nu \Psi -{1\over 2}g_{\mu\nu}
        \nabla^\alpha \Psi \nabla_\alpha\Psi\right)
        \label{1:ST}  \ . 
\end{eqnarray}
The right hand side of this Eq.~(\ref{1:ST}) should be identified with 
$-E_{\mu\nu}-\lambda g_{\mu\nu}$.
 Hence, the  condition  $E^\mu{}_\mu =0$ becomes
\begin{eqnarray}
\Box\Psi = - {\omega \over 3\Psi} 
        \nabla^\mu \Psi \nabla_\mu \Psi  
        - {4\over 3} \left( \Lambda  - \lambda \Psi \right)  \ .
	\label{1:KG1}
\end{eqnarray}
This is the equation for the radion $\Psi$. However, we also
have the equation for $\Psi$ from the action (\ref{1:action}) as 
\begin{eqnarray}
\Box \Psi = \left( {1\over 2\Psi} - { \omega' \over 2\omega} \right)
	\nabla^\alpha \Psi \nabla_\alpha \Psi  
	-{\Psi \over 2\omega} R + {\Psi \over \omega} \Lambda'    \ ,
	\label{1:KG2}
\end{eqnarray}
where the prime denotes the derivative with respect to $\Psi$. 
In order for these two Eqs.~(\ref{1:KG1}) and (\ref{1:KG2}) to be compatible, 
$\Lambda$ and $\omega$ must satisfy 
\begin{eqnarray}
&&-{\omega\over 3\Psi}
	={1\over 2\Psi}-{ \omega' \over 2\omega}
	\label{eq1}
	\ ,  \\
&&{4\over 3} \left( \Lambda-\lambda\Psi\right)
	={\Psi\over\omega}\left( 2\lambda -\Lambda'\right)
	\label{eq2}  \ ,
\end{eqnarray}
where we used  $R= 4\lambda$ which comes from the trace part of 
Eq.~(\ref{1:SMS}). Eqs.~(\ref{eq1}) and (\ref{eq2}) can be integrated as  
\begin{eqnarray}
   \Lambda (\Psi) = \lambda + \lambda \gamma \left( 1-\Psi \right)^2   \ , \quad
  \omega (\Psi ) = {3\over 2} {\Psi \over 1-\Psi}  \ ,  
\end{eqnarray}
where the constant of integration $\gamma$ represents the ratio
 of the cosmological constant on the negative tension brane to that on 
 the positive tension brane. 
 Here, one of constants of integration is absorbed by rescaling of $\Psi$.
 In doing so, we have assumed the constant of integration is positive.
 We can also describe the negative tension brane if we take the 
 negative signature.

Thus, we get the effective action 
\begin{eqnarray}
S_{\rm eff}
	&=&\int d^4 x \sqrt{-g} \left[ {1\over 2} \Psi R 
	-{3 \over 4( 1-\Psi )} \nabla^\mu \Psi \nabla_\mu \Psi 
	\right. \nonumber\\
	&&\left.\qquad\qquad\qquad
	- \lambda - \lambda \gamma (1-\Psi)^2 \right] \ .
\end{eqnarray}
Surprisingly, this completely agrees with the previous 
result (\ref{A:action}). Our simple symmetry 
principle $E^\mu{}_\mu =0$ has determined the action completely. 
 
 As we have shown in \cite{KSS}, if $\gamma <-1$
 there exists a static deSitter two-brane solution
 which turns out to be unstable. In particular,  
 two inflating branes  can collide at $\Psi = 0$. 
 This process is completely smooth for the observer on the brane.  
 This fact led us to the born-again scenario.
 The similar process occurs also in the ekpyrotic (cyclic) model~\cite{turok}
 where the moduli approximation is used. It can be shown that the moduli
 approximation is nothing but the lowest order truncation of the low energy
 gradient expansion method developed by us~\cite{KS}. Hence,
 it is of great interest to see the leading order corrections due
 to KK modes to this process.  

  Let us apply the procedure explained above 
 to the higher order case:
\begin{eqnarray}
S_{\rm eff}
\!\!&=&\!\!
	{1\over 2} \int d^4 x \sqrt{-g} \left[ \Psi R - 2\Lambda (\Psi )
	-{\omega (\Psi) \over \Psi} \nabla^\mu \Psi \nabla_\mu \Psi \right]
        \nonumber\\
&&\!\!
	+\int d^4 x \sqrt{-g} \left[
	A(\Psi) \left( \nabla^\mu \Psi \nabla_\mu \Psi \right)^2
	+B(\Psi) \left( \Box \Psi \right)^2  \right. \nonumber\\
&&\left. \quad\qquad	+C(\Psi)\nabla^\mu \Psi \nabla_\mu \Psi \Box \Psi
	+D(\Psi) R~\Box \Psi 
	      \right. \nonumber\\
&&\left. \quad\qquad
	+ E(\Psi) R \nabla^\mu \Psi \nabla_\mu \Psi
        + F(\Psi) R^{\mu\nu} \nabla_\mu \Psi \nabla_\nu \Psi  
              \right. \nonumber\\
&&\left. \quad\qquad
        + G(\Psi) R^2     
	+ H(\Psi) R^{\mu\nu} R_{\mu\nu} 
	        \right. \nonumber\\
&&\left. \quad\qquad
	+I(\Psi) R^{\mu\nu\lambda\rho} R_{\mu\nu\lambda\rho} 
	+\cdots  \right]  \ .
	\label{setup}
\end{eqnarray}
   Now we impose the conformal symmetry  on the fourth order derivative terms
  in the action (\ref{setup}) as we did in the  previous example. 
  Starting from the action (\ref{setup}), one can read off the equation for 
  the metric from which $E_{\mu\nu}$ can be identified. 
  The compatibility between the equations of motion for $\Psi$
  and the equation $E^\mu{}_\mu =0$ determines the coefficient functionals
  in the action (\ref{setup}).
 
Thus, we find the 4-dimensional effective action with KK corrections as
\begin{eqnarray}
S_{\rm eff} 
	&=& \int d^4 x \sqrt{-g} \left[ {1\over 2} \Psi R 
	-{3 \over 4( 1-\Psi )} \nabla^\mu \Psi \nabla_\mu \Psi 
        \right. \nonumber\\
&&\left.\quad
	-\lambda-\lambda\gamma (1-\Psi)^2
	\right]\nonumber\\
&&	+\ell^2 \int d^4 x \sqrt{-g}
	\left[
	{1 \over 4 (1-\Psi)^4} \left( \nabla^\mu \Psi \nabla_\mu \Psi \right)^2
	\right. \nonumber\\
&&\left.
	+{1\over (1-\Psi)^2} \left( \Box \Psi \right)^2 
	+{1\over (1-\Psi)^3} \nabla^\mu \Psi \nabla_\mu \Psi\Box \Psi
	\right. \nonumber \\
&&\left.
	+ {2\over 3(1-\Psi)} R \Box \Psi 
	+{1\over 3(1-\Psi)^2}  R \nabla^\mu \Psi \nabla_\mu \Psi 
	\right. \nonumber\\
&&\left.
	+jR^2+ k R^{\mu\nu} R_{\mu\nu}  \right] \ ,
	\label{action}
\end{eqnarray}
where constants $j$ and $k$ can be interpreted as 
 the variety of the effects of the bulk gravitational waves.
It should be noted that this action becomes non-local after
 integrating out the radion field. This fits the fact that
 KK effects are non-local usually.  
  In principle, we can continue this calculation to any order of 
 derivatives.

\section{Conclusion}

We have developed the low energy gradient expansion scheme
 to give  insights into the physics of the braneworld
 such as the black hole physics and the cosmology.
 In particular, we have concentrated on the specific questions 
 in this paper.
 Here, we summarize our answers obtained by the gradient expansion method.
 Our understanding of RS braneworlds would be useful for other brane models.

\subsection{Single-brane model}

\noindent
{\bf Is the Einstein theory  recovered even in the non-linear regime?}\\

 We have obtained the effective theory at the lowest order
as
\begin{equation}
\overset{(4)}{G}{}^\mu{}_{\nu}  
	={\kappa^2\over\ell}T^\mu{}_{\nu} 
	-{2\over \ell} \chi^{\mu}{}_{\nu} \ .
\end{equation}
Here we have the correction $\chi_{\mu\nu}$ 
which can be interpreted as the dark radiation 
in the cosmological situation. 

 On the other hand, in the linearized gravity, 
 the conventional Einstein theory is recovered at
 low energy. This is because the out-going boundary condition is imposed.
 In other words, the asymptotic AdS boundary condition is imposed.
 In the nonlinear case, this corresponds to the requirement that
  the dark radiation term $\chi_{\mu\nu}$ must be zero. 
 For this boundary condition, the conventional Einstein theory is recovered.
 Hence, the standard Friedmann equation holds. 
 
 In this sense, the answer is yes.

\vskip 0.5cm
\noindent
{\bf How does the AdS/CFT come into the braneworld?}\\

The CFT emerges as the constant of integration which satisfies
the trace anomaly relation
\begin{equation}
\tau^\mu{}_{\mu}  
	=-{1\over 8}\left( R^\alpha{}_{\beta}R^\beta{}_{\alpha} 
	-{1\over 3}R^2\right)-\beta\Box R      \ .
\end{equation}
This constant can not be determined a priori. 
 Here, the AdS/CFT correspondence could come into the braneworld.
 Namely, if we identify some CFT with $\tau_{\mu\nu}$, then we can
 determine the boundary condition.

\vskip 0.5cm
\noindent
{\bf How are the AdS/CFT and geometrical approach  related?}\\

The key quantity in the geometric approach is obtained as
\begin{equation}
\overset{(2)}{E}{}^{\mu}{}_{\nu} 
	=\ell^2\left[P^\mu{}_{\nu}+2\tau^\mu{}_{\nu} 
	+2\alpha{\cal S}^\mu{}_{\nu} 
	+{2\over 3}\beta{\cal K}^\mu{}_{\nu}\right]     \ .
\end{equation}
The above expression contains $\tau_{\mu\nu}$ which can be interpreted
 as the CFT matter. Hence, once we know $E_{\mu\nu}$, no enigma remains.
 In particular, $P_{\mu\nu}\approx \pi_{\mu\nu}$ is independent
 of the $\tau_{\mu\nu}$. In odd dimensions, there exists
 no trace anomaly, but $P_{\mu\nu}$ exists. In 4-dimensions, 
 $\pi^\mu{}_\mu$ accidentally coincides with the trace anomaly in CFT. 
 
 It is interesting to note that the high energy and 
 the Weyl term corrections found in the geometrical
    approach merge into the CFT matter correction found in the 
    AdS/CFT approach.

\subsection{Two-brane model}

\noindent
{\bf How is the geometrical approach consistent with the Brans-Dicke
 picture?}\\

 In the geometrical approach, no radion seems to appear. On the other
 hand, the linear theory predicts the radion as the crucial quantity.
 The resolution can be attained by obtaining $E_{\mu\nu}$
 ($\chi_{\mu\nu}$ in our notation). The resultant expression
\begin{eqnarray}
\chi^{\mu}{}_{\nu}&=&-{\kappa^2(1-\Psi)\over 2 \Psi} 
	\left( \overset{\oplus}{T}{}^{\mu}{}_{\nu} 
	+(1-\Psi)\overset{\ominus}{T}{}^{\mu}{}_{\nu}\right)  \nonumber\\
&&	-{\ell\over 2\Psi} \left[ \left(  \Psi^{|\mu}{}_{|\nu} 
	-\delta^\mu_\nu  \Psi^{|\alpha}{}_{|\alpha} \right) 
	\right. \nonumber \\
&&	\left.+{3 \over 2(1 -\Psi )} \left( \Psi^{|\mu}  \Psi_{|\nu}
  	-{1\over 2} \delta^\mu_\nu  \Psi^{|\alpha} \Psi_{|\alpha} 
  	\right) \right]   \nonumber
\end{eqnarray}
contains the radion in an intriguing way. 
The dark radiation consists of the radion and the matter. 

 We have shown that the radion transforms the Einstein theory with
  Weyl correction into the conformally coupled scalar-tensor theory 
  where the radion plays the role of the scalar field.

\vskip 0.5cm
\noindent
{\bf What replaces  the AdS/CFT correspondence 
in the two-brane model?}\\

 In the case of the single-brane model, the out-going boundary condition 
 at the Cauchy horizon is assumed. This conforms to AdS/CFT correspondence.
 Indeed, the continuum KK-spectrum are projected on the brane as CFT matter.
 
 On the other hand, the boundary condition in the two-brane system 
 allows only the discrete KK-spectrum. Hence, we can not expect CFT
 matter on the brane. Instead, the radion controls the bulk/brane
 correspondence in two-brane model. 
 In fact, the higher derivative terms of the radion mimics
 the effect of the bulk geometry (KK-effect) as we have shown explicitly. 

 Hence, the AdS/CFT correspondence does not exist. 
 Instead, the AdS/radion correspondence exists.
 
\begin{acknowledgements}
This work was supported in part by  Grant-in-Aid for  Scientific
Research Fund of the Ministry of Education, Science and Culture of Japan 
 No. 155476 (SK) and  No.14540258 (JS) and also
  by a Grant-in-Aid for the 21st Century COE ``Center for
  Diversity and Universality in Physics".  
\end{acknowledgements}


\end{document}